\documentclass[a4paper,11pt]{amsart}

\usepackage{graphicx}

\newcommand{\resolution}{_jpg}
\newcommand{\myinclude}[1]{\includegraphics[width=12.5cm,angle=0,trim=0 0 0 0]
                                                {#1\resolution}}
\newcommand{\new}[1]{{#1}}
\newcommand{\old}[1]{}

\begin{document}

\title{Significance of log-periodic signatures in cumulative noise}

\author{Hans-Christian Graf v. Bothmer}
\address{Institut f\"ur Mathematik (C), Universit\"at Hannover, Welfengarten 1, D-30167 Hannover}
\email{bothmer@math.uni-hannover.de}
\date{\today}

\begin{abstract}
Using methods introduced by Scargle we derive a cumulative version of 
the Lomb periodogram that exhibits frequency independent statistics when applied to
cumulative noise. We show how this cumulative Lomb periodogram allows us to estimate
the significance of log-periodic signatures in the S\&P 500 anti-bubble that started in August 2000.
\end{abstract}
	
\maketitle

\section{Introduction}

\new{
Speculative bubbles, crashes and depressions are some of the most puzzling phenomena in 
financial markets. Even though these are rare events they occur much more often, than expected from standard models of financial markets. For example the otherwise very successfull GARCH-Model predicts only one crash of a magnitute comparable to the events of 1929 and 1987 in 16000 years \cite{outliers}. This suggest that the very largest crashes are outliers that are caused by phenomena not considered in the standard models. 

Several authors \cite{precursors}, \cite{FeigenbaumFreud}, \cite{vandewalle} have argued that this behavior is natural, if one considers the stock-market as a complex system. When this system is  far away from critical points, standard models are a good description of the systems dynamics, but when a critical point is approached the dynamics change completely.
In this picture a crash occurs, when a critical point is reached.

The theory of critical phenomena predicts that all observables of a complex system near a critical point are scale invariant. In the case of stock markets
this would mean that the price $p(t)$ near a critical point should follow a power-law
\[
               \log p(t) \sim A + B(t_c-t)^\beta.
\]
where $t_c$ is the time of crash. Unfortunately it is rather difficult to distinguish this from
the exponential growth
\[
               \log p(t) \sim A + Bt
\]
predicted by standard theory. 

Help in this situation comes from the concept of discrete scale invariance. If the system considered has a hirachical structure it might scale only in accordance with these hirachies. This allows the scaling coefficient $\beta$ to be a complex number and the observables will obey
\begin{align*}
              \log p(t) \sim &Re(A + B(t_c-t)^\beta) \\
                                     &= 
              A + B(t_c-t)^\alpha + C(t_c-t)^\alpha \cos (\omega ln(t_c-t) + \phi),
\end{align*}
Such a power-law with log-periodic signature is much easier to detect than a pure power-law. In fact sizable log-periodic signatures
have been detected before many financial crashes, including the ones from 1929 and 1987 (\cite{precursors}, \cite{FeigenbaumFreud}, \cite{vandewalle}, \cite{CriticalCrashes}, \cite{nasdaq}).
Also they have been found after the bursting of some speculative bubbles, for example gold after 1980, the Nikkei-Index after 1990 and most recently after the bursting of the new economy
bubble in 2000 (\cite{antibubbles}, \cite{How_much_longer}).

Still there is some reason for doubt. As several authors have pointed out (\cite{No_log_after_earthquakes}, \cite{Feigenbaum_Kritik_1}), log-periodic signatures 
can often arise by chance in systems that do not exhibit discrete scale invariance. Even simple Brownian motion produces log-periodic signatures quite often. Therefore there has been
a heated discussion about the significance of the found log-periodic} 
\old{In the last years there has been a heated discussion about the presence of log-periodic}
signatures
in financial data \cite{Feigenbaum_Kritik_1}, \cite{Feigenbaum_Antwort}, \cite{Feigenbaum_Kritik_2}. 

One tool to detect these signatures has been the Lomb periodogram
introduced by Lomb \cite{Lomb} and improved by Scargle in \cite{Scargle}. An important property of
Scargle's periodogram is that the individual Lomb powers of independently normal distributed noise
follow approximately an exponential distribution.  

Unfortunately this property is lost, when
one calculates Scargle's periodogram for cumulative noise, where the differences between
two observations are independently normal distributed. In this Brownian-Motion case, the expected Lomb powers are much greater for small frequencies than for large ones. Therefore the
significance of large Lomb powers at small frequencies is difficult to estimate. Huang, Saleur, Sornette and Zhou tackle this problem and several related ones with extensive Monte Carlo simulations in \cite{No_log_after_earthquakes} and \cite{Heavy_tailed}. 

\old{Here} \new{In this paper} we present an analytic approach \new{to the problem of estimating the significance of log-periodic signatures}.  
\old{In the first part of this paper we introduce}\new{We start by introducing} a small correction to Scargle's Lomb periodogram, \old{to 
make}\new{that makes} the distribution of Lomb powers exactly exponential for independently normal distributed noise.

In the second part we use the same methods to derive a normalisation of the Lomb periodogram
that assures an frequency independent exponential distribution of Lomb powers for cumulative noise. 

In the last section
we apply these new methods to estimate the significance of log-periodic signatures in so called S\&P 500-anti-bubble after the crash of 2000.  We show how our methods greatly simplify the whole
analysis and derive that there is about a $6\%$ chance that a signature like the one detected by Sornette and Zhou in \cite{How_much_longer} arises by chance
if one only considers frequencies smaller than $10.0$. 
If one searches all frequencies up to the Nyquist frequency, peaks of this height become
much more common. 

Furthermore we detect equally significant peaks at 
harmonics of the fundamental frequency of Sornette and
Zhou. This complements evidence for a more sophisticated modelling of the S\&P 500 anti-bubble
by Sornette and Zhou in \cite{Renormalization}.

We would like to thank Volker M\"oller for hosting our database of financial data and donating
cpu-time for the Monte Carlo simulations that lead to this paper.

\section{Independent Noise}

\newcommand{\sumj}[1]{\sum_{j=1}^{N_0}  #1}
\newcommand{\sumsin}{\sumj{ X_j \sin \omega t_j}}
\newcommand{\sumcos}{\sumj{ X_j \cos \omega t_j}}
\newcommand{\sumsinsq}{\sumj{\sin^2 \omega t_j}}
\newcommand{\sumcossq}{\sumj{\cos^2 \omega t_j}}
\newcommand{\sumcossin}{\sumj{\cos \omega t_j \sin \omega t_j}}

\newcommand{\psquare}[1]{\Bigl( #1 \Bigr)^2}                                                
                                               
Consider the classical periodogram
\[
    P(\omega) = \frac{1}{N_0} 
                           \left[ 
                               \psquare{\sumcos} + \psquare{\sumsin}
                           \right].
\]
This is a generalization of the Fourier transform to the case of unevenly spaced 
measurements. Unfortunately this classical form has difficult statistical behavior 
for uneven spacing. Scargle therefore proposed in \cite{Scargle} a normalized from of 
the periodogram, that restores good statistical properties in the case where
all $X_j$ are independently normal distributed with mean zero and
variance $\sigma_0^2$.  For this he observes that
\[
      S(w) = \sumcos
\]
and
\[
      C(w) = \sumsin
\]
are again normally distributed, with variances
\[
     \sigma_c^2 =   N_0 \sigma_0^2 \sumcossq 
\]
and
 \[
     \sigma_s^2 =   N_0 \sigma_0^2 \sumsinsq. 
\]
Also he claims that for normally distributed random variables with unit variance
their sum of squares is exponentially distributed. Therefore he proposes to 
look at
\[                            
    P(\omega) = 
                           \left[ 
                               \frac{\psquare{\sumcos}}{\sigma_c} + 
                               \frac{\psquare{\sumsin}}{\sigma_s}
                           \right].
\]
\new{Scargles claim, that this sum is exponentially distributed is only correct}
\old{Strictly speaking this is only correct} when $S(\omega)$ and $C(\omega)$ are independent. This is
approximately true, when the observation times $t_j$ are "not to badly bunched". In other
cases whole variance/covariance matrix
\[
   \Sigma = \begin{pmatrix}
                       \sigma_c^2  & \sigma_{cs} \\
                       \sigma_{cs} & \sigma_s^2
                     \end{pmatrix}
\]
with 
 \[
     \sigma_{cs}=   N_0 \sigma_0^2 \sumcossin
\]
has to be considered.

\old{Then the values of a quadratic form
\[
      \bigl(C(\omega), S(\omega)\bigr)
      Q
      \begin{pmatrix} C(\omega)\\ S(\omega) \end{pmatrix}
\]
are exponentially distributed if 
\[
       Q \Sigma Q = Q
\]
or equivalently $Q = \Sigma^{-1}$,  if $\Sigma$ is invertible. In our case we have}

\new{If $\Sigma$ is invertible the values of the quadratic form
\[
      P(\omega) = \frac{1}{2} \bigl(C(\omega), S(\omega)\bigr)
      \Sigma^{-1}
      \begin{pmatrix} C(\omega)\\ S(\omega) \end{pmatrix}
\]
are exponentially distributed, i.e.
\[
       Probability(P(\omega) > z )= \exp(-z).
\]
by standard facts about multivariate normal distributions 
(for example \cite[Theorem 1.4.1]{Muirhead}).  Explicitly we can calculate}                 
\[
      \Sigma^{-1} = \frac{1}{\sigma_c^2\sigma_s^2-\sigma_{cs}^2}
                               \begin{pmatrix}
                                    \sigma_s^2 & -\sigma_{cs} \\
                                    -\sigma_{cs} & \sigma_c^2
                               \end{pmatrix}.
\]
This means the natural form of the periodogram is
\[
     P(\omega) = \frac{1}{2} 
                            \frac{\sigma_s^2 C^2(\omega) + \sigma_c^2 S^2(\omega)  
                                     - 2\sigma_{cs} C(\omega) S(\omega) }
                                    {\sigma_c^2\sigma_s^2-\sigma_{cs}^2},
\]
which reduces to Scargle's case for $C$ and $S$ independent ($\sigma_{cs}=0$), and to the classical
case for even spacing ($\sigma_c=\sigma_s$). This $P(\omega)$ is always exponentially distributed. 

As Scargle we also replace $t_j$ by $t_j - \tau$ with 
\[
      \tau = \frac{1}{2\omega}\arctan \frac{\sumj{\sin 2\omega t_j}}{\sumj{\cos 2\omega t_j}}
\]
in all formulas, to make the diagram time-translation invariant.

\section {Cumulative Noise}

\newcommand{\sumy}{\sum_{k=1}^j Y_k}

\begin{figure}[ht]
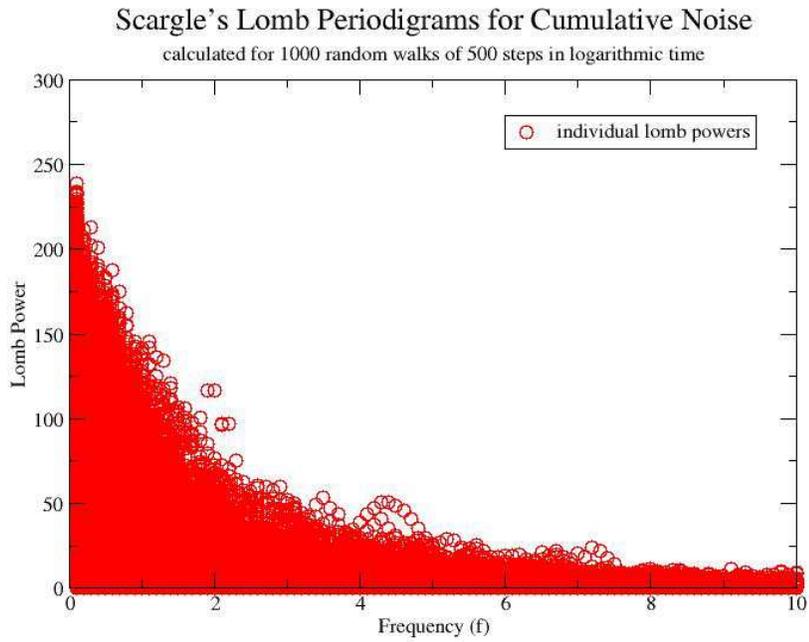

\begin{center}
\myinclude{figure1}
\caption{Scargle's periodogram for 1000 random walks with normal distributed
                innovations of zero mean
                and unit variance. Each random walk has 500 steps which are assumed to
                occur at times $\log(1) \dots \log(500)$}
\label{random_all_nodetrend}
\end{center}                
\end{figure}

Assume now, that the differences $Y_j =(X_j - X_{j-1})$ are independently normal distributed with
zero mean and variance $\sigma_0^2$. Then the distribution of powers in Scargle's periodogram
depends on the angular frequency $\omega = 2\pi f$. See figure \ref{random_all_nodetrend} for the result of a Monte Carlo simulation
of this case. In small frequencies much higher Lomb powers can occur by chance, then in
high frequencies. This reflects the well known fact that the expected value at frequency $f$ 
of the evenly spaced Fourier transform of a brownian motion 
is proportional to $1/f^2$. Our idea is now to 
normalize the periodogram by adjusting Scargle's methods to this case.

\begin{figure}[ht]
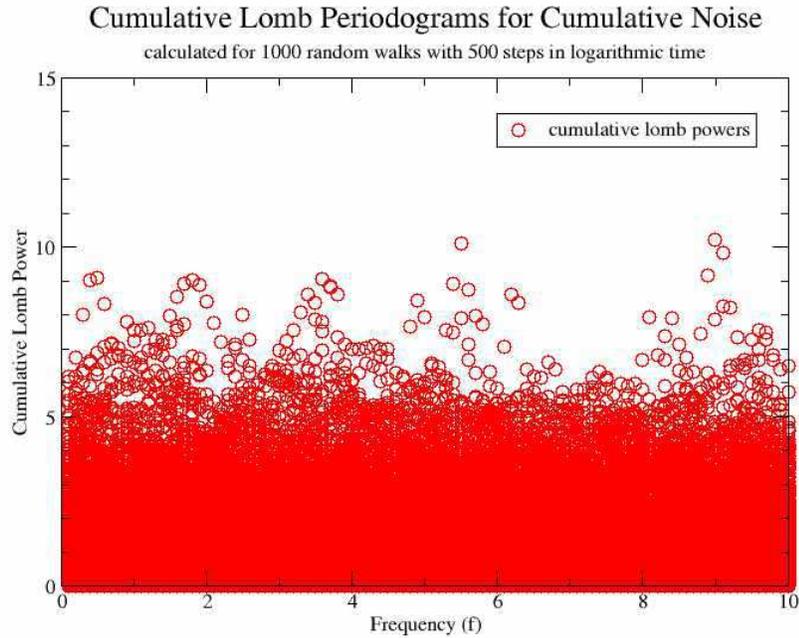

\myinclude{figure2}
\caption{Cumulative Lomb periodogram for 1000 random walks with innovations of zero mean
                and unit variance. Each random walk has 500 steps which are assumed to
                occur at times $\log(1)\dots\log(500)$. For the normalisation $\sigma_0=1$ has
                been used.}
\label{random_all_cumulative}
\end{figure}

First notice that  
\[
        X_j - X_0 = \sumy
\]
in this situation. From this we obtain
\begin{align*}
     C(\omega) &= \sum_{j=1}^n (X_j-X_0) \cos \omega t_j \\
                         &= \sum_{j=1}^n \Bigl(\sum_{k=1}^j Y_k\Bigr) \cos \omega t_j \\
                         &= \sum_{k=1}^n Y_k \sum_{j=k}^{n}  \cos \omega t_j 
\end{align*}
and a similar formula for $S(\omega)$. Notice that $C(\omega) $ and $S(\omega) $ are still normally
distributed with 
\begin{align*}
          \sigma_c^2 &= \langle C^2(\omega)\rangle\\
                               &= \sum_{kl} \langle Y_k, Y_l \rangle \Bigl(\sum_{j=k}^{n}  \cos \omega t_j\Bigr)
                                                                                                       \Bigl(\sum_{j=l}^{n}  \cos \omega t_j\Bigr) \\
                                &= \sigma_0^2  \sum_{k=1}^n \Bigl(\sum_{j=k}^{n}  \cos \omega t_j\Bigr)^2
\end{align*}
and
\begin{align*}
          \sigma_s^2 &= \langle S^2(\omega)\rangle\\
                               &= \sum_{kl} \langle Y_k, Y_l \rangle \Bigl(\sum_{j=k}^{n}  \sin \omega t_j\Bigr)
                                                                                                       \Bigl(\sum_{j=l}^{n}  \sin \omega t_j\Bigr) \\
                                &= \sigma_0^2  \sum_{k=1}^n \Bigl(\sum_{j=k}^{n}  \sin \omega t_j\Bigr)^2
\end{align*}
and
\begin{align*}
          \sigma_{cs}&= \langle C(\omega)S(\omega) \rangle\\
                               &= \sum_{kl} \langle Y_k, Y_l \rangle \Bigl(\sum_{j=k}^{n}  \sin \omega t_j\Bigr)
                                                                                                       \Bigl(\sum_{j=l}^{n}  \cos \omega t_j\Bigr) \\
                                &= \sigma_0^2  \sum_{k=1}^n \Bigl(\sum_{j=k}^{n}  \cos \omega t_j\Bigr)
                                                                                       \Bigl(\sum_{j=k}^{n}  \sin \omega t_j\Bigr).
\end{align*}
With these values
\[
     P(\omega) = \frac{1}{2} 
                            \frac{\sigma_s^2 C^2(\omega) + \sigma_c^2 S^2(\omega) 
                                                                                    - 2\sigma_{cs} C(\omega)S(\omega)}
                                    {\sigma_c^2\sigma_s^2-\sigma_{cs}^2},
\]
is again exponentially distributed:
\[
       Probability\bigl(P(\omega) > z\bigr) = \exp(-z)
\]
In particular the distribution is now independent of $\omega$ as exemplified by figure \ref{random_all_cumulative} which shows the
cumulative Lomb periodogram for a Monte Carlo simulation as above.

\begin{figure}[ht]
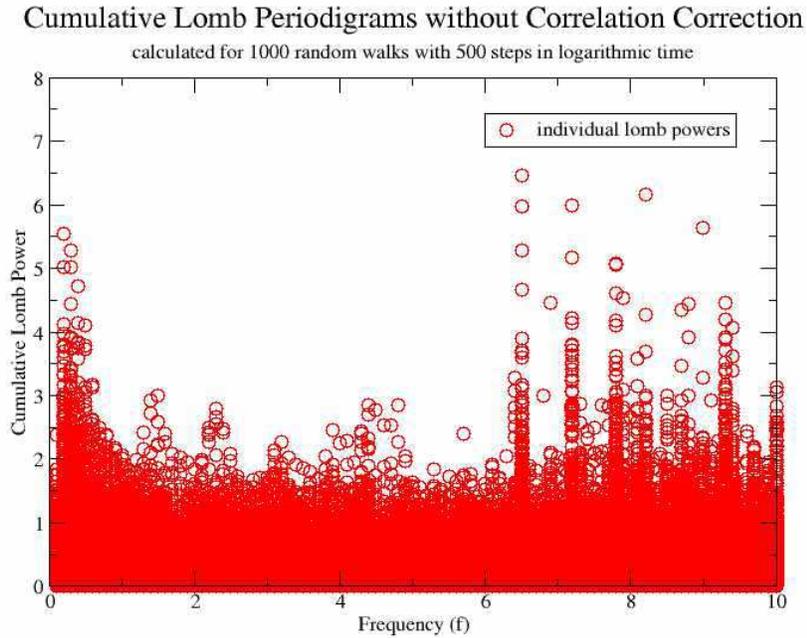

\myinclude{figure3}
\caption{Cumulative Lomb periodograms without the correction for correlation between 
                $C(\omega)$ and $S(\omega)$ of 1000 random walks with innovations of zero mean
                and unit variance. Each of the random walks has 500 steps which are assumed to
                occur at times $\log(1)\dots\log(500)$. $\sigma_0$ was estimated
                for each dataset. The ommission of the correction terms introduces spurious peaks
                at several frequencies.}
\label{random_all_nocorrection}
\end{figure}

Notice that is essential to consider the correlation between $C(\omega) $ and $S(\omega) $ 
in this case. Figure
\ref{random_all_nocorrection} shows the cumulative Lomb periodograms for 1000 random walks of length 500 in logarithmic time
without the correlation term above. Notice how this introduces spurious peaks at several
frequencies.

Again we replace $t_j$ by $t_j - \tau$ with 
\[
      \tau = \frac{1}{2\omega}\arctan \frac{\sumj{\sin 2\omega t_j}}{\sumj{\cos 2\omega t_j}}
\]
in all formulas, to make the diagram time-translation invariant.

\section{Application to log-periodicity in financial data}

Recently Sornette and Zhou have suggested that there is a log periodic signature
in the S\&P 500 index after the bursting of the new economy bubble \cite{How_much_longer}. 
Among other methods they use Scargle's
periodogram with $\tilde{t}_j = \log(t_j-t_c)$ and $X_j$ the logarithm of the index price $j$ days
after a critical date $t_c$ \new{for frequencies $0\le f \le 10$}. To account
for a nonlinear trend of the form
\[
        \quad A + B(t_j-t_c)^\alpha
\]
they first detrend the index values. For this they use a value of $A$
obtained from a nonlinear fit of  
\[
        (*) \quad \quad A + B(t_j-t_c)^\alpha + C (t_j-t_c)^\alpha\cos(\omega \ln(t_j-t_c) + \phi)
\]
to the logarithm of the index price 
and then determine $B$ and $\alpha$ for different choices of $t_c$ by linear regression.

\new{
We follow the same procedure, but estimate $A$ by a different method. For this we consider
the correlation coefficient
\[
          r_A = \frac{\sum_i (\tilde{t}_j - \bar{t})(\log(X_j-A)-\bar{X})}
                             {\sqrt{\sum_i (\tilde{t}_j - \bar{t})^2}\sqrt{\sum_i(\log(X_j-A)-\bar{X})^2}}
\]
with $\bar{t}$ the average of the $\tilde{t_j}$'s and $\bar{X}$ the average
of the $\log(X_j-A)$'s. This correlation coefficient measures how well $\log(X_j-A)$
can be approximated by a linear model
\[
       \log B + \alpha \log(t_j - t_c).
\]
Notice that $r_A$ does not depend on $B$ and $\alpha$. 
Since correlation coefficiens near $+1$ and $-1$ indicate strong explanatory value of the
proposed trend, while coefficients near $0$ indicate that a trend with this value of $A$ seems not to
be present, we first search the value of $A$ that maximises $r_A^2$. Then, like Sornette and Zhou, we find $B$ and $\alpha$ by linear regression}

\new{
With this method we address a critique of Feigenbaum \cite{Feigenbaum_Kritik_2}. He points out that the detection of log-periodicity by lomb periodograms is not independent of the detection via nonlinear fits of equation $(*)$ to the price data, if values produced by the second are reused in the first.  Our approach eliminates this dependance. }

\old{
We eliminate the dependence to the nonlinear fitting procedure by using a slightly different method:
Fixing a critical date $t_c$ we optimise $A$ for the best correlation between logarithmic time
$\tilde{t}_j$ and  $\log(X_j-A)$. 
}

\begin{figure}[ht]
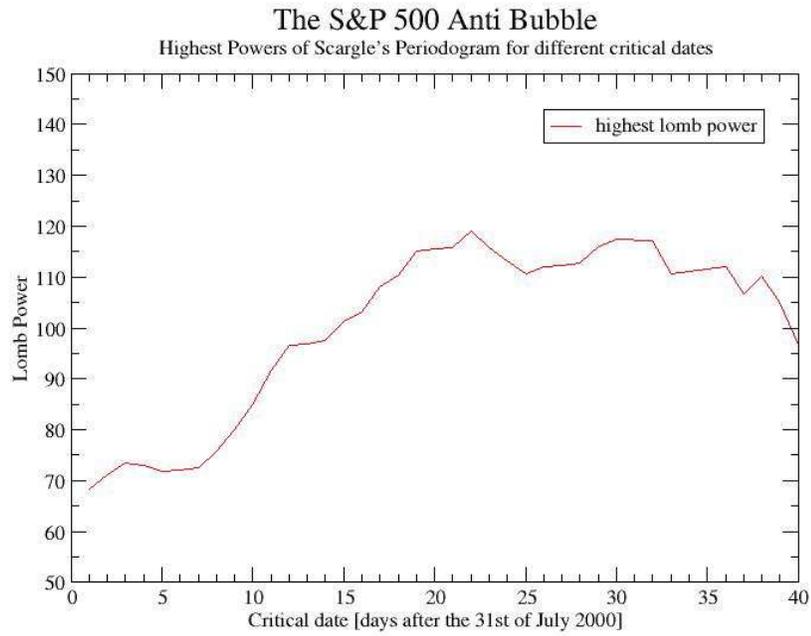

\myinclude{figure4}
\caption{Highest peaks in Scargle's periodograms of 2-year windows of S\&P 500 data
               starting from different days in August and September 2000. Before calculating the
               periodogram the price data has
               been detrended according to the procedure described in the text.}               
\label{snp500_highest}
\end{figure}

Figure \ref{snp500_highest} shows the highest Lomb powers obtained by this method for $2$-year intervals starting
from August $1$st to September $5$th, 2000. The highest peak in our dataset is observed for
the critical date August $22$th, 2000. This is in reasonable agreement with the critical date found by
Sornette and Zhou (August $9$th, 2000). 

\begin{figure}[ht]
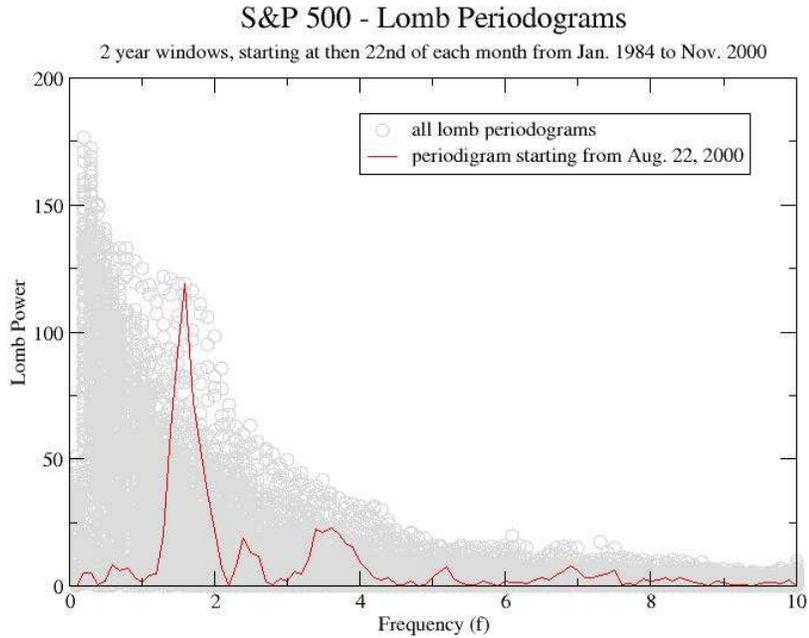

\myinclude{figure5}
\caption{202 Scargle periodograms for 2-year windows of S\&P 500 data
               starting from 22nd of each month. Before calculating the
               periodogram the price data has
               been detrended according to the procedure described in the text.
               For one window the detrending procedure did not converge}               
\label{snp500_allfreq}
\end{figure}

To estimate the significance of this peak we calculated
the 203 Lomb periodograms of $2$-year intervals starting at the $22$nd of each month from Jannuary $1984$ to November $2000$. 
Figure \ref{snp500_allfreq} shows a comparison of these periodograms with the one from August $22$nd, 2002.
Notice that the distribution of Lomb-powers depends strongly on the frequency. For small frequencies high Lomb powers are much more probable than at high frequencies. The absence of large powers
for very small frequencies is due to the detrending procedure. These facts have also been observed 
by \cite{No_log_after_earthquakes} in a somewhat different setting.

Notice also that the peak at $f\approx 1.6$ is nevertheless quite large, in fact it is the largest one observed at this frequency. But how probable is it that we have a peak of this relative
height for {\sl  any} of the tested frequencies? The SnP500 data set is not large enough, to 
estimate this probability accurately, but a count shows that there are 39 datasets that have
at least one frequency with a peak that is higher than any of the other at the same frequency.
This gives a naive estimate of $39/202 \approx 19,3 \%$.

\begin{figure}[ht]
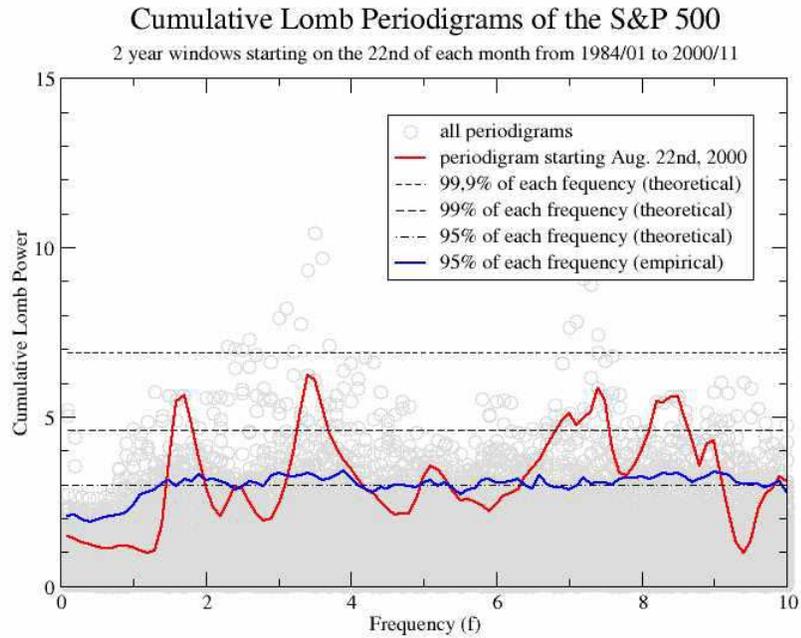

\myinclude{figure6}
\caption{203 cumulative periodograms for 2-year windows of S\&P 500 data
               starting from 22nd of each month. The price data has not been detrended.
               $\sigma^2_0 = 0.000119403$ has been estimated from the total $18$ years of data.}               
\label{snp500_cumulative}
\end{figure}

The cumulative Lomb periodogram proposed above improves and greatly simplifies this analysis. Figure \ref{snp500_cumulative}
shows the cumulative Lomb periodograms
of the S\&P 500 anti-bubble together with the cumulative Lomb
periodograms of all other datasets. A detrending of the data as above was not necessary.
Notice that there are now several peaks
of comparable size at frequencies $1.7$, $3.4$, $7.4$ and $8.4$. The fact that these lie close to
the harmonics of $1.7$ compares well to the results of \cite{Renormalization}.

We have included the theoretical $99.9\%$, $99\%$ and $95\%$-quantiles for each frequency derived above together with the $95\%$-quantiles of the actual S\&P 500 data. Notice the excellent agreement for frequencies greater than $1$. 

\begin{figure}[ht]
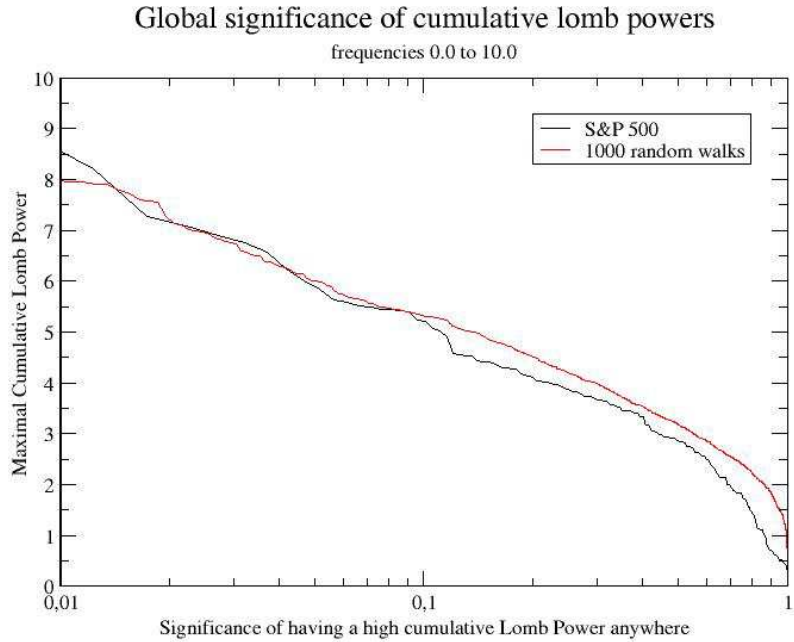

\myinclude{figure7}
\caption{Highest peaks of cumulative periodograms for 1000 random walks compared with
               those of 203 cumulative periodograms for 2-year windows of S\&P 500 data
               starting from 22nd of each month. The price data has not been detrended.
               $\sigma^2_0 = 0.000119403$ has been estimated from the total $18$ years of data.}               
\label{snp500_significance}
\end{figure}

Since the height of
the peaks is now largely independent from the frequency, we can estimate the global significance of the peak at $1.7$ with cumulative Lomb power $5.61$ by counting the number of cumulative Lomb-periodograms with at least
one peak of higher power \new{for frequencies $0 \le f \le 10$}. We find $12$ of those which implies a significance of approximately
$12/203 \approx 5.9\%$. In figure \ref{snp500_significance} we compare the global significance of peaks in cumulative Lomb periodograms of the S\&P 500 with those obtained from a Monto Carlo simulation of
brownian motion \new{with zero mean and unit variance} . There is a reasonable agreement for significances smaller than $0.1$ \new{even though brownian motion is only a very
rough model of stock market prices.}

Notice that the \old{this result} \new{global significance of a peak} depends strongly on the number of frequencies considered. If we
consider all frequencies up to the Nyquist frequency for the average sampling interval
\[
      f_N = \frac{N_0}{2T} = \frac{500}{2\log(500)} \approx 40.2,
\]
a peak of height $5.61$ as above is found in $27.6\%$ of \old{the} \new{simulated} periodograms \old{(estimated again by a Monte Carlo simulation of 1000 random walks with 500 steps each).}
\new{and in $28.3\%$ of the historical S\&P 500 periodograms. In this light the choice of
$0 \le f \le 10$ by Sornette and Zhou seems to call for an apriori justification if one wants to keep
up the claim of a significant log-periodic signature around $f=1.7$.
}    
\old{So it seems that one needs some a priori reason for considering only small frequencies, to justify the claim of a significant log-periodic signature in this case.}

\section{Conclusion}

We have indrocuced a new version of the Lomb periodigram that exhibits good statistical
properties when applied to cumulative noise. With this we were able to detect the log-periodic
signature in the S\&P 500 anti-bubble with better significance than with the ususal periodigram,
even without a detrending procedure. More importantly this method allows us to estimate the significance of the found log periodic signature which is a reasonable $5.9\%$ if we only consider fequencies smaller than $10.0$, and a disappointing $27.6\%$ if one considers all frequencies
up to the Nyquist frequency for the average sampling interval. We also detect cumulative
Lomb peaks of similar significance at harmonics of the fundamental frequency $1.7$.


\begin{thebibliography}{VBMA98}

\bibitem[AJ99]{antibubbles}
Didier~Sornette Anders~Johansen.
\newblock Financial ``anti-bubbles'': Log-periodicity in gold and nikkei
  collapses.
\newblock {\em International Journal of Modern Physics C}, 10(4):563--575,
  1999.

\bibitem[AJ00]{nasdaq}
Didier~Sornette Anders~Johansen.
\newblock The nasdaq crash of april 2000: Yet another example of
  log-periodicity in a speculative bubble ending in a crash.
\newblock {\em European Physical Journal B}, 17:319--328, 2000.

\bibitem[DS96]{precursors}
Jean-Philippe~Bouchaud Didier~Sornette, Anders~Johansen.
\newblock Stock market crashes, precursors and replicas.
\newblock {\em J.Phys.I France}, 6:167--175, 1996.

\bibitem[Fei01a]{Feigenbaum_Kritik_2}
James~A. Feigenbaum.
\newblock More on a statistical analysis of log-periodic precursors to
  financial crashes.
\newblock {\em Quantitative Finance}, 1(5):527--532, 2001.

\bibitem[Fei01b]{Feigenbaum_Kritik_1}
James~A. Feigenbaum.
\newblock A statistical analysis of log-periodic precursors to financial
  crashes.
\newblock {\em Quantitative Finance}, 1(3):346--360, 2001.

\bibitem[HJL{\etalchar{+}}00]{No_log_after_earthquakes}
Y.~Huang, A.~Johansen, M.~W. Lee, H.~Saleur, and D.~Sornette.
\newblock Artifactual log-periodicity in finite-size data: Relevance for
  earthquake aftershocks.
\newblock {\em J. Geophysical Research}, 105(B12):28111--28123, 2000.

\bibitem[JAF96]{FeigenbaumFreud}
P.G.O.~Freud James A.~Feigenbaum.
\newblock Discrete scale invariance in stock markets before crashes.
\newblock {\em Int. J. Mod. Phys.}, 10:3737--3745, 1996.

\bibitem[JS98]{outliers}
A.~Johansen and D.~Sornette.
\newblock Stock market crashes are outliers.
\newblock {\em European Physical Journal B}, 1:141--143, 1998.

\bibitem[Lom76]{Lomb}
N.~R. Lomb.
\newblock Least-squares frequency analysis of unequally spaced data.
\newblock {\em Astrophysics and Science}, 39:447--462, 1976.

\bibitem[Mui82]{Muirhead}
Robb~J. Muirhead.
\newblock {\em Aspects of multivariate statistical theory}.
\newblock Wiley series in probability and mathematical statitstics. John Wiley
  \& Sons, 1982.

\bibitem[Sca82]{Scargle}
Jeffrey.D. Scargle.
\newblock Studies in astronomical time series analysis. {II}. {S}tatistical
  aspects of spectral analysis of unevenly spaced data.
\newblock {\em Astrophysical Journal}, 263:835--853, 1982.

\bibitem[SJ01]{Feigenbaum_Antwort}
D.~Sornette and A.~Johansen.
\newblock Significance of log-periodic precursors to financial crashes.
\newblock {\em Quantitative Finance}, 1(4):452--471, 2001.

\bibitem[SLJ99]{CriticalCrashes}
D.~Sornette, O.~Ledoit, and A.~Johansen.
\newblock Critical crashes.
\newblock {\em Risk Magazine}, 12(1):91--95, 1999.

\bibitem[SZ02]{How_much_longer}
D.~Sornette and W.-X. Zhou.
\newblock The {US} 2000-2002 market descent: How much longer and deeper?
\newblock {\em Quantitative Finance}, 2(6):468--481, 2002.

\bibitem[VBMA98]{vandewalle}
N.~Vandewalle, Ph. Boveroux, A.~Minguet, and M.~Ausloos.
\newblock The krach of october 1987 seen as a phase transition: amplitude and
  universality.
\newblock {\em Physica A}, 255(1-2):201--210, 1998.

\bibitem[ZS02]{Heavy_tailed}
Wei-Xing Zhou and Didier Sornette.
\newblock Statistical significance of periodicity and log-periodicity with
  heavy-tailed correlated noise.
\newblock {\em Int. J. Mod. Phys. C}, 13(2):137--170, 2002.

\bibitem[ZS03]{Renormalization}
Wei-Xing Zhou and Didier Sornette.
\newblock Renormalization group analysis of the 2000-2002 anti-bubble in the
  {US} {S}\&{P} 500 index: Explanation of the hierarchy of 5 crashes and
  prediction.
\newblock {\em Physica A}, 2003.

\end{thebibliography}

\newcommand{\etalchar}[1]{$^{#1}$}

\pagebreak


\pagebreak

\pagebreak

\pagebreak

\pagebreak

\pagebreak 

\pagebreak

\end{document}